\documentclass{emulateapj}

\newcommand{\halpha}{H$\alpha$}
\newcommand{\vlsr}{\ensuremath{v_{\mathrm{LSR}}}}
\newcommand{\kms}{\ensuremath{\mathrm{km\,s}^{-1}}}

\newcommand{\paper}{\emph{Letter}}
\newcommand{\pip}{PIP}

\newcommand{\hmss}[4]{#1$^{\mathrm h}$ #2$^{\mathrm m}$ #3$\fs$#4}
\newcommand{\dms}[3]{#1$\arcdeg$ #2$\arcmin$ #3$\arcsec$}

\newcommand{\knotg}{Knot g}

\newcommand{\gzz}{G00}

\shorttitle{SUBARU HDS Observation of Tycho}
\shortauthors{Lee et al.}

\begin{document}

\title{SUBARU HDS Observations of a Balmer-Dominated Shock in Tycho's 
Supernova Remnant
\footnote{Based on data collected at Subaru Telescope, which is
operated by the National Astronomical Observatory of Japan}
}

\author{Jae-Joon Lee,\altaffilmark{1,2} Bon-Chul Koo,\altaffilmark{1}
John Raymond,\altaffilmark{3} Parviz Ghavamian,\altaffilmark{4}
Tae-Soo Pyo,\altaffilmark{5} Akito Tajitsu,\altaffilmark{5} and Masahiko Hayashi\altaffilmark{5}}
\altaffiltext{1}{Astronomy Program, Department of Physics and Astronomy,
  Seoul National University, Seoul 151-742, Korea }

\altaffiltext{2}{jjlee@astro.snu.ac.kr}

\altaffiltext{3}{Harvard-Smithsonian Center for Astrophysics, 60 Garden Street, Cambridge, MA 02138}

\altaffiltext{4}{Department of Physics and Astronomy, Johns Hopkins University, 3400 North Charles Street, Baltimore, MD 21218}

\altaffiltext{5}{Subaru Telescope, National Astronomical Observatory of Japan, 
650 North A'oh$\bar{\mathrm{o}}$k$\bar{\mathrm{u}}$ Place, Hilo, HI 96720}

\begin{abstract}
We present an \halpha\ spectral observation of a Balmer-dominated 
shock on the eastern side of Tycho's supernova remnant using SUBARU Telescope. 
Utilizing the High Dispersion Spectrograph (HDS), we measure the spatial
variation of the line profile between preshock and postshock gas.
Our observation clearly shows a broadening and centroid shift of the 
narrow-component postshock \halpha\ line relative to the \halpha\,
emission from the preshock gas. 
The observation supports the existence of a thin precursor
where gas is heated and accelerated ahead of the shock.
Furthermore, the spatial profile of the emission ahead of the Balmer filament
shows a gradual gradient in the \halpha\ intensity and line width
ahead of the shock.  We propose that this region ($\sim\, 10^{16}$ cm)
is likely to be the spatially resolved precursor.  The line width increases from
$\sim 30\ \kms$ up to $\sim 45\ \kms$
and its central velocity shows a redshift of $\sim5\ \kms$ across the shock front.
The characteristics of the precursor are consistent with a cosmic
ray precursor, although a possibility of a fast neutral precursor
is not ruled out.
\end{abstract}

\keywords{ISM:supernova remnants -- ISM: individual (Tycho,
  G120.1+1.4) -- Shock Waves -- line: profiles}

\section{INTRODUCTION}
\label{sec:intro}

Balmer-dominated filaments are the signature of non-radiative
shocks propagating into partially neutral medium %
\citep{1978ApJ...225L..27C}. 
The \halpha\ line profile is composed of two distinctive
components (narrow and broad)  
representing the velocity distribution of preshock and postshock
gases, respectively \citep{1980ApJ...235..186C}. 
High resolution spectroscopic observations of several of these
shocks have revealed that the width of the narrow component is 
unusually large (30 $\sim$ 50 \kms) for ambient neutral hydrogen,
and it was proposed that the gas was heated 
in a precursor thin enough ($\lesssim 10^{17}$ cm) to avoid complete ionization of
hydrogen \citep[see][and references therein]{2001ApJ...547..995G,2003A&A...407..249S}.%
\footnote{We refer to this precursor explicitly as a
``thin precursor'', to avoid potential confusion with 
a photoionization precursor which will be shortly
introduced.}
Two likely candidates are cosmic-ray (CR) and fast neutral precursors
\citep{1994ApJ...420..286S,1994ApJ...420..721H,2001ApJ...547..995G}.
Both scenarios predict significant Doppler shifts of preshock gas.
No clear indication of such a shift of the \halpha\ narrow component 
is reported \citep[but see][]{2004ApJ...605L.113L}.
A careful comparison of postshock and preshock line profiles 
with high spectral resolution is crucial for
confirming the existence of such a precursor.

The blast wave of Tycho's SNR, the historical remnant of the 1572
supernova, has been known to exhibit 
Balmer-dominated emission 
\citep{1971ApJ...168...37V
,1978ApJ...224..851K
}.
The region of the brightest \halpha\ emission 
\citep[Knot g from][]{1978ApJ...224..851K}
is located along the north-eastern edge of the remnant. 
Faint, diffuse optical emission extends $\sim$1 pc
ahead of the Balmer-dominated filaments
\citep[][\gzz\ hereafter]{2000ApJ...535..266G}.  
This feature has been identified as a photoionization precursor (\pip)
produced by photoionization of the preshock gas by \ion{He}{2} $\lambda
304$\AA\ emission from behind the blast wave
(\gzz).
The estimated temperature of the \pip\ 
  is $\sim 1.2 \times 10^4$ K, which is not
high enough to explain the observed width of the Balmer narrow component.

The existence of \halpha\ emission from the PIP makes Tycho an unique
target for the study of nature of the thin precursor, as we can investigate
the change of the line profile across this precursor by comparing the
line emission from the PIP (preshock) and that of \knotg\ (postshock).
In this \paper, we present 
high resolution (Echelle) long slit \halpha\ spectra  of
Tycho \knotg\  and its PIP using the
SUBARU  High Dispersion Spectrograph \citep[][]{2002PASJ...54..855N}.
Our observations reveal the line broadening and the Doppler shift of the 
the narrow component, 
providing strong evidence for the existence of a cosmic ray or fast
neutral precursor. 

\section{OBSERVATIONS AND RESULTS}
\label{sec:obs}

\begin{figure}
\epsscale{1.20}
\plotone{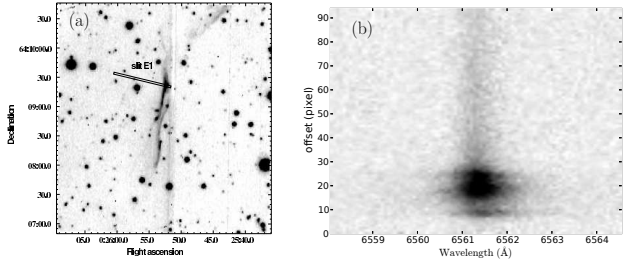}
\caption{
(a) \halpha\ image of Tycho \knotg\  with position of the 2\arcsec\,$\times$60\arcsec\, slit
overlaid.
(b) The fully-processed \halpha\ 2-d spectrum (only the western part
of the slit is shown). The bright emission knot at the bottom is \knotg.
\label{fig:one}}
\end{figure}

The spectrosopic observation was carried out on 2004 October 1.
Longslit Echelle spectroscopy of the \halpha\ line was performed 
using order-blocking filters (HDS standard setup
``stdHa''). This gives a spectral coverage of $6540$ \AA $\sim$
$6690$ \AA\ over the $60\arcsec$ of slit length. 
The slit was centered at 
$\alpha(2000)$, $\delta(2000)$ = (\hmss{00}{25}{56}{5}, \dms{64}{09}{28}) with
position-angle of  $76\arcdeg$ (measured E of N), covering both \knotg\  and
its \pip\ simultaneously (Fig.~\ref{fig:one}(a)). 
A total of 3$\times$30 minutes of source exposure was obtained, and 
the same amount of
exposure for nearby sky. 
The spectrum was binned by 2 along the slit direction and 4 along the
dispersion direction before the readout. 
The pixel scale after the binning is 0.27\arcsec\ pixel$^{-1}$ 
($9.3\times 10^{15}$ cm pixel$^{-1}$ at a
distance of 2.3 kpc \citep{1978ApJ...224..851K}) and 
$0.08$ \AA\, 
respectively.
The slit width was $2 \arcsec$, which gives velocity
resolution of $17$ \kms.
The seeing was $0\farcs 5 \pm 0\farcs 1$. 
The processing of the SUBARU data included a typical CCD
preprocessing 
(including overscan correction and flat fielding) 
and two-dimensional spectral extraction.
A wavelength calibration solution is obtained from the spectrum of a Th-Ar
lamp. 
The source spectrum was sky-subtracted, and normalized 
using the spectra of standard stars. 
The uncertainty in the wavelength calibration is estimated to be
around $0.2$ \kms\ at the wavelength of \halpha. 

\begin{figure}
\epsscale{.9}
\plotone{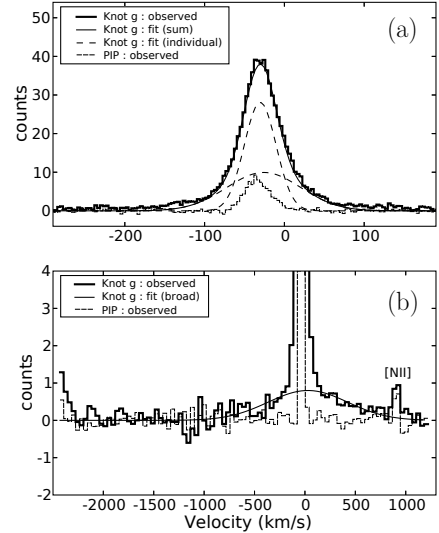}
\caption{
(a) \halpha\ spectrum of \knotg\  and the
PIP. The spectrum of \knotg\  is fitted with three Gaussian components, 
and two narrowest components are displayed (thin dashed) together
with their sum (thin solid).
(b) Same as (a) except the full observed velocity range is presented and
the spectra are binned. The \knotg\  broad component from the above fit
is displayed as a thin solid line.
\label{fig:spec}
}
\end{figure}

In Fig.~1(b), we present the fully processed two dimensional
spectrum of \halpha\ line. 
The bright patch at the bottom with three distinct 
local emission peaks corresponds to \knotg, and 
the faint emission extending to the top of the image
corresponds to the PIP.
The average \halpha\ spectrum of \knotg\ and that of PIP are
shown together in  Fig.~\ref{fig:spec}.
The velocity width of the \knotg\ narrow component line is clearly larger than that of the
PIP \halpha\ line.  In addition, the velocity centroid of the \knotg\ narrow component is
slightly redshifted ($5.5 \pm 0.6\ \kms$) relative to that of the PIP \halpha\ line. And as clearly seen in Fig.~\ref{fig:spec}(b),
the \knotg\  spectrum shows a very broad ($\sim$ 1,000 \kms), faint
\halpha\ line.
The small peak near $900$ \kms\ is the [\ion{N}{2}] $\lambda 6583.4$~\AA\
line from the PIP.

The spectrum of the PIP is well fitted by a single Gaussian, and yields a FWHM
of $33.8\pm 0.8$ \kms\ and centroid velocity of $-35.8\pm 0.6$ 
\kms\ (in LSR frame).\footnote{throughout this paper, we give line widths 
corrected for instrumental broadening and velocity in LSR frame.}
The measured FWHM of [\ion{N}{2}] $\lambda
6583.4$~\AA\  in the PIP is $\sim 23$ \kms.
If the broadening were purely thermal, the widths of the lines whould
be inversely proportional to the square root of their atomic masses.
The expected width of the [\ion{N}{2}] line in this case is 
$1/\sqrt{14}=0.267$ of \halpha. As this is significantly narrower than 
what is observed, a significant amount of nonthermal broadening
is suggested.
If we simply assume that the observed line widths are a
convolution of thermal and nonthermal  broadenings, 
the estimated thermal temperature is $\sim 13,000$ K,
consistent with $12,000$ K of \gzz. 
It is also possible that the large line width is due to a residual of
Galactic \halpha\ emission.

An adequate fit to the \knotg\ spectrum requires three Gaussian components.
They have velocity widths of $45.3\pm 0.9$ \kms\ (narrow), 
$108\pm 4$ \kms\ (intermediate) and $931 \pm 55$ \kms\ (broad),
with central velocities of $-30.3\pm 0.2$, $-25.8\pm 0.8 $ and
$29 \pm 18$ \kms, respectively.
This result confirms that the \halpha\ narrow component line of \knotg\  is
redshifted and broadened relative to that of the PIP. 
The broad component (FWHM$\sim 1,000\ \kms$) should correspond to the
previously reported $\sim 2,000\ \kms$ component
\citep{2001ApJ...547..995G}.  The relatively
narrower width of our observation is probably due to the insensitivity
of our spectroscopic configuration to the very broad line, although the
possibility of temporal variation (e.g., by crossing a density jump) does exist.
The abrupt density increase might be possible if the shock is
propagating into the edge outskirts of the dense cloud
\citep{2004ApJ...605L.113L}.
The existence of the intermediate width component has already been
reported by \gzz. 
It may be produced when slow protons picked up by the postshock magnetic
field undergo a secondary charge exchange. 
Alternatively, it might be an artifact of the assumption of Gaussian
distributions, which would not necessarily be appropriate if the
motions are non-thermal.

The characteristics  of the narrow component \halpha\ line
of \knotg\ are consistent with \gzz, except that the velocity centroid of the
line in our data is significantly different from theirs
($\vlsr = -30.3\pm 0.2$ from our data vs.\ $-53.9\pm 1.3\ \kms$ from \gzz).
We have carefully assessed our wavelength calibration
, but couldn't find any significant source of error. 
The wavelengths of night sky lines observed near \halpha\ matched well with
those of VLT UVES night-sky emission line catalog
\citep{2003A&A...407.1157H}
within error of $0.02$ \AA\ ($\sim 1\ \kms$). 
The \halpha\ spectrum of blank sky is also consistent with 
nearby WHAM spectra \citep{2003ApJS..149..405H}.
Furthermore, the central velocity of [\ion{N}{2}] $\lambda$6583.4 \AA\ 
emission line from the \pip\ is consistent with that of \halpha.
Although we believe that our wavelength calibration solution
is quite secure, the discrepancy with the previous observation should be
confirmed by an independent observation.
For the rest of this paper, we mainly concentrate on
the line width and the relative variation of central velocity.

\section{Location of the Shock Front and Discovery of a Thin ($10^{16}$ cm)  Precursor}

\begin{figure}
\epsscale{.95}
\plotone{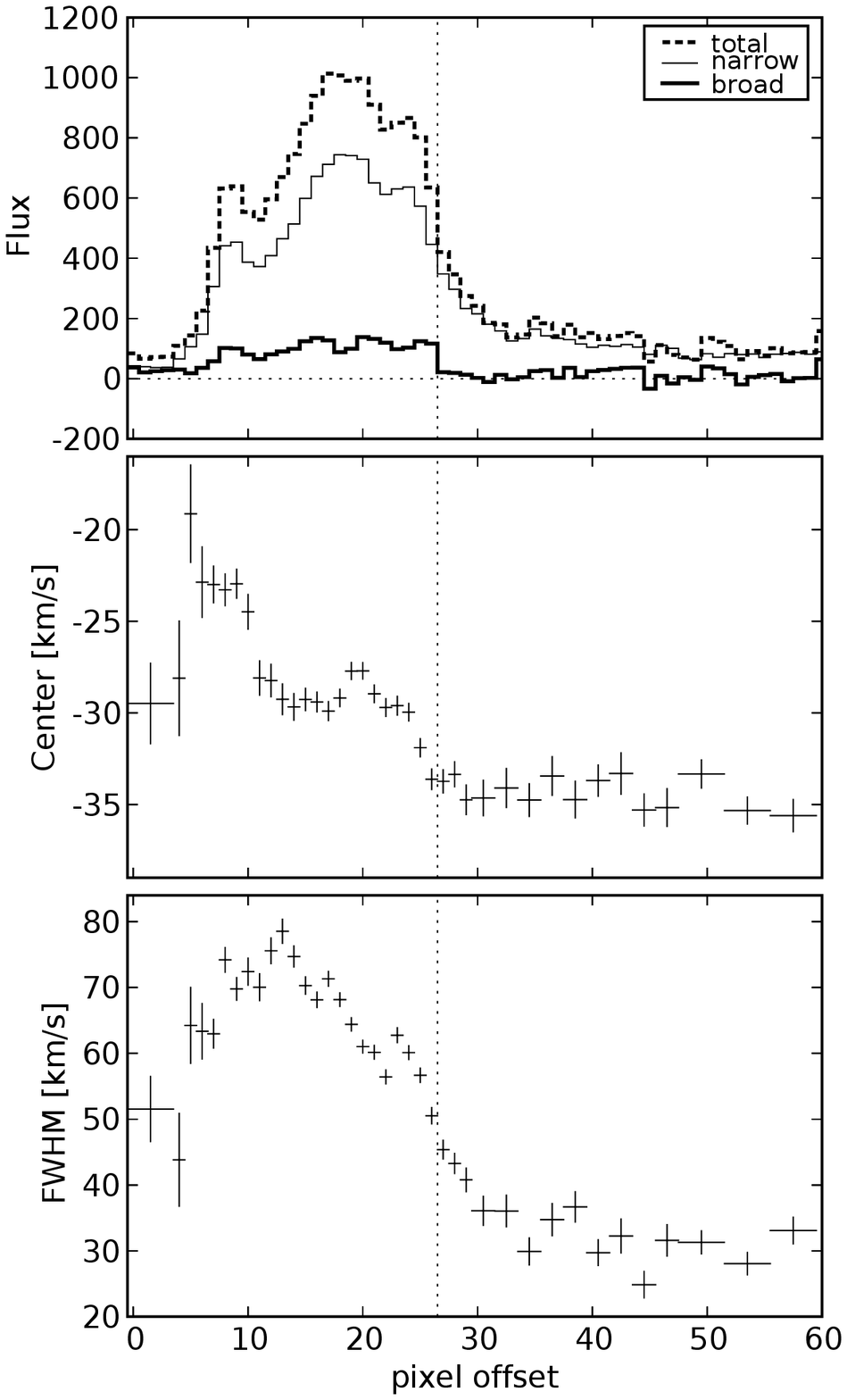}
\caption{
(top) Integrated fluxes along the slit position of \halpha\
spectrum for a given velocity range.
The thin solid line shows emission in the range $-60 < \vlsr < 10\ \kms$,
representative of \halpha\ narrow component,  
while the thick solid line shows emission in the range 
$-400 < \vlsr < -120\ \kms$ and
$+100 < \vlsr < +350\ \kms$, for the broad component. 
The thick dashed line shows the summed intensity for both components.
($-400 < \vlsr < +350\ \kms$). Middle and bottom panels: central
velocities and FWHMs from fits with a single Gaussian. 
The proposed location of shock front is marked as a vertical dotted line.
\label{fig:two}
}
\end{figure}

Fig.~\ref{fig:two} shows the spatial variation of the \halpha\ line
profile along the slit. 
In the top panel, we separately plot the integrated intensities of
the representative \halpha\ of narrow and broad components, 
together with their sum
(refer to the figure caption for the velocity range of each component).
In the middle and bottom panels, we plot the central
velocities and FWHMs at each slit position from the {\em single}
Gaussian fit to the full profile.  
In all three panels, the x-axis represents the pixel offset from the
southwest end of the slit. 
Although three Gaussian components are actually required to adequately fit the 
\halpha\, profile of \knotg, the non-uniqueness of the
multiple-component fitting can complicate the interpretation of the 
different emission line components.  Therefore, we plot the fit result
from a single Gaussian and describe our interpretation of this fit 
in the region of \knotg\ as following:
First, the central velocities closely 
match those of the narrow component.
Fitting with multiple  components 
leaves the fitted line centroids unchanged to within the errors.
Second, the FWHMs 
basically trace the spatial variation of the narrow 
component width, but the presence 
of the intermediate and the broad components contribute significantly to the width.
When fit with three Gaussian components, the large widths of 
the \halpha\ narrow component in \knotg\ from the single profile 
models ($60 \sim 80\ \kms$)  are reduced to
$40 \sim 50\ \kms$.

Inspection of Fig.~\ref{fig:two} reveals that there 
exists only one location (pixel offset 27, marked as a dotted vertical
line) where the intensity and width of the \halpha\ line exhibits an abrupt jump.
The jump is most noticeable for the broad component, where 
the flux is virtually zero toward the direction of ambient medium 
(i.e., toward positive pixel offsets). 
The central velocity of the narrow component also shows rapid change
around this location. 
The FWHM also steeply increases behind this point, but this 
might be largely
due to 
the sudden appearance of the broad component.  
The fact that the intensity of the
broad component, which is associated with the postshock gas,
shows a significant jump at this location suggests that 
it corresponds to the location of the shock front.

The \halpha\ intensity in the PIP region is nearly constant along the
slit, but shows an rapid 
rise just before the shock front, i.e., from pixel
offset 32 to 27 in Fig.~\ref{fig:two}. 
The intensity of the \halpha\ line in the PIP region is about 10\%
of the observed peak value in \knotg, and it reaches about half 
the peak very near the shock. 
The line width also increases 
from 30 to 45 \kms\ within this region.
We propose that the steep increase of the \halpha\ intensity
accompanied with line broadening corresponds to 
a thin precursor where gas is heated and accelerated.  
Unlike the line broadening, the observed velocity centroids shift 
slightly behind the shock instead of the precursor region. 
This does not necessarily indicate that the broadening and the Doppler
shift takes at different region as this 
is likely due to geometrical projection effects. 
The \halpha\ intensity profile gives 
an e-folding thickness 
of $1.4 \pm 0.4$ pixel (after accounting for seeing),
which corresponds to $(1.4 \pm 0.4) \times 10^{16}$ cm at a distance
of 2.3 kpc. 

The observed emission of Knot g shows a few local peaks indicating 
possible substructure, e.g., a collection of shock
tangencies seen in projection along the line of sight. 
This leads to a possibility that the `precursor' is simply the results
of geometric projection of fainter Balmer-dominated filament lying just
ahead of Knot g.
If we assume that this filament has a line profile similar to that of
\knotg, i.e., have a same broad-to-narrow ratio, 
then a detectable amount of broad component is expected. 
The flux profile of broad component plotted in
Fig.~\ref{fig:two} would show similar gradual increase in
the precursor region, which is not seen. 
Also, no hint of a broad component is found in a summed spectra of 
precursor region, which is expected to show up.
%
%
Examining the archival 
{\em Chandra} observation of Tycho's SNR
(the analysis is presented by Warren et al. 2005), we did not find evidence
of X-ray emission extending ahead of Knot g.  Therefore, there is no
strong supporting evidence in favor of a projection, although
we cannot rule out the possibility. 

\section{Nature of the Thin Precursor}
\label{sec:vel-struct}

\label{cr-precursor}

As the shock is nearly tangential to our line of sight, 
the actual amount of bulk gas acceleration could be much larger than
the observed Doppler shift.  
The shock angle can be inferred from the radial velocity shift of
the broad component of \knotg\ relative to the narrow component
\citep{1980ApJ...235..186C}, which is
measured to be $\sim 60\ \kms$.  
As our observation could be insensitive to this broad component,
using this value is rather conservative.
On the other hand, \citet{2001ApJ...547..995G} reported a redshift of
$\sim 130\ \kms$, 
but their field of view is different from ours.
We take these two values as limits and give shock angle of
$2\arcdeg-5\arcdeg$ assuming a shock velocity of 2,000 \kms.
We consider $5\ \kms$ to be the representative 
redshift of the narrow component compared to unperturbed medium, 
which gives actual acceleration of $60\sim130\ \kms$.

The line width of the narrow component at the shock
front ($\sim 45\,\kms$) corresponds to
a neutral hydrogen temperature of $40,000$ K, if the line broadening is 
purely thermal, or lower, if there is a significant non-thermal broadening
such as a wave motion.
Neutral hydrogen atoms and protons may have similar velocity
distributions due to their charge exchange interactions. 
But that of electrons can be different, which would depend on the
heating mechanism in the precursor.
%
%
In the following, we estimate the electron temperature in the 
thin precursor (TP)
%
from the observed intensity increase of factor 5.
Since the observed spectrum is an 
integrated emission along the line of sight where
a significant contribution from PIP region is expected, the actual
emissivity increase within the precursor should be much greater.
We assume that regions of 
the PIP and the TP
 are represented by two concentric shells with thicknesses of
$\delta_{\mathrm{PIP}} \simeq 10^{16}$\,cm and
$\delta_{\mathrm{TP}} \simeq 1$\,pc (\gzz), respectively,
and that both have an inner diameter of the Tycho
(5.4\,pc). Then the ratio of path length through each shell along the
tangential direction of the shock front is $\simeq 18$.
This implies that the emissivity increase
in the TP could be as large as a factor of $90$. 
This value should be regarded as an upper limit as it is likely that the
filament is nearly flat and tangent to the line of sight.
%
%
The collisional excitation rate of \halpha\ at
$\mathrm{T}_{\mathrm{e}} \sim 40,000$\,K is $2,000$ times greater than
that of $12,000$\,K which is the temperature of
the PIP region (\gzz). This value greatly exceeds the emissivity
increase of $90$, and implies a few possibilities. 
The electron temperature can be less than $40,000$\,K, either if 
%
observed \halpha\ line width has significant nonthermal broadening,
or if $\mathrm{T}_{\mathrm{e}}$ is intrinsically less than 
$\mathrm{T}_{\mathrm{p}}$. 
On the other hand, the estimated emissivity increase might be
explained if 
the emissivity increase due to high $\mathrm{T}_{\mathrm{e}}$
is suppressed  by ionization of neutrals. 
Since the
existence of Blamer-dominated filaments requires a significant fraction
of neutral hydrogen atoms to survive within precursor, this possibility is
less favored.

The two likely  candidates for this
precursor are fast neutral and CR precursors.  
The momentum and energy carried by upstreaming fast neutrals can be large
enough to explain the observed heating and acceleration in the precursor,
but the available model calculations 
do not predict significant heating by these neutrals
\citep{1996MNRAS.280..103L,2005PhDT..........K}. 
Furthermore, It is difficult to reproduce the small range
of FWHMs observed for 
narrow component \halpha\ lines from shocks of different velocities 
\citep{1994ApJ...420..286S,1994ApJ...420..721H}.
The observed characteristics of the precursor are consistent 
with a CR precursor. CR acceleration in the shock does require a
precursor \citep[e.g.,][]{1987PhR...154....1B} and there has been growing
evidence of CR acceleration in young SNRs including Tycho itself
\citep[e.g.,][]{2005ApJ...634..376W}. 
Significant heating and acceleration are expected to
happen in the CR precursor \citep{1987PhR...154....1B}.
Interaction of CR particles in
the upstream generates Alf\'{v}en waves and significant amplification
of magnetic field has been suggested \citep{2001MNRAS.321..433B}. 
Although a measurement of the preshock magnetic field is hardly available,
a high value ($40\,\mu$G) of preshock magnetic field is
suggested for Tycho 
\citep{2002A&A...396..649V}, 
which may explain the line width of the narrow component.

To conclude, our SUBARU observation clearly
has shown  the line broadening and the Doppler shift 
between the preshock gas and postshock gas.
This strongly suggests the existence of 
the thin precursor. 
Furthermore, the precursor itself is likely to be resolved.
Given the lack of observational constraints on CR ion acceleration in SNR
shocks, the tentatively measured width ($\sim 10^{16}$ cm) of the thin
precursor whose primary candidate is the CR precursor is promising.
Follow-up observation, such as a high resolution imaging with {\em HST},
would clearly 
resolve the structures of the precursor.

\acknowledgments{We thank to the anonymous referee for valuable 
comments. 
This work was supported by the 
Korea Research Foundation (grant No. R14-2002-058-01003-0).
JJL has been supported in part by the BK 21 program.}



\end{document}